\documentclass{article}

\pdfoutput=1 

\UseRawInputEncoding
\usepackage{subfigure}
\usepackage{arxiv}
\usepackage[square,numbers,sort&compress]{natbib}
\usepackage[utf8]{inputenc} 
\usepackage[T1]{fontenc}    
\usepackage{hyperref}       
\usepackage{url}            
\usepackage{booktabs}       
\usepackage{amsfonts}       
\usepackage{nicefrac}       
\usepackage{microtype}      
\usepackage{lipsum}		
\usepackage{graphicx}
\usepackage{doi}

\title{An Adaptable IoT Rule Engine Framework for Dataflow Monitoring and Control Strategies}

\date{} 					

\author{ 
{\hspace{1mm}Ken Chen} \\
	School of Software and Microelectronics, Peking University\\
	\texttt{aken5930@outlook.com} \\
}

\renewcommand{\headeright}{}
\renewcommand{\undertitle}{}

\hypersetup{
pdftitle={An Adaptable IoT Rule Engine Framework for Dataflow Monitoring and Control Strategies},
pdfsubject={q-bio.NC, q-bio.QM},
pdfauthor={Ken Chen},
pdfkeywords={IoT, Rule Engine, Dataflow Monitoring and Control},
}
\begin{document}
\maketitle

\begin{abstract}
The monitoring of data generated by a large number of devices in Internet of Things (IoT) systems is an important and complex issue. Several studies have explored the use of generic rule engine, primarily based on the RETE algorithm, for monitoring the flow of device data. In order to solve the performance problem of the RETE algorithm in IoT scenarios, some studies have also proposed improved RETE algorithms. However, implementing modifications to the general rule engine remains challenges in practical applications. The Thingsboard open-source platform introduces an IoT-specific rule engine that does not rely on the RETE algorithm. Its interactive mode attracted attention from developers and researchers. However, the close integration between its rule module and the platform, as well as the difficulty in formulating rules for multiple devices, limits its flexibility. This paper presents an adaptable and user-friendly rule engine framework for monitoring and control IoT device data flows. The framework is easily extensible and allows for the formulation of rules contain multiple devices. We designed a Domain-Specific Language (DSL) for rule description. A prototype system of this framework was implemented to verify the validity of theoretical method. The framework has potential to be adaptable to a wide range of IoT scenarios and is especially effective in where real-time control demands are not as strict.
\end{abstract}


\section{Introduction}
The extensive data generated by Internet of Things (IoT) devices, along with the diverse needs of users, highlights the necessity for customized monitoring of device data flows. In the past few years, researchers predominantly relied on general rule engines, primarily based on the RETE\cite{forgy1989rete} algorithm, to address these challenges\cite{el2017sre,kiran2014adaptive,mainetti2015novel,mazon2018rules,luo2021scalable,cambra2017iot}. Rule engine is a sort of software that separates rule definitions from their implementation. Typically, rule consists of IF-THEN statements, where the action in the THEN statement is executed if the condition in the IF statement is met. Rule engine interprets these rule statements and integrates them into the software system. The RETE algorithm provides an efficient method for matching rules with data. It is currently the most widely used approach for constructing rule engines.

The RETE algorithm optimizes the calculation of matching results by caching intermediate operation results, thereby reducing the time complexity of rule matching. This allows for the reuse of previous calculation results if a certain input attribute remains unchanged in the next batch of data. However, in IoT scenarios, device messages change rapidly, resulting in a low reuse rate of intermediate results. Thus, the space-for-time strategy often proves inadequate. To address this issue, researchers have proposed improved versions of the RETE algorithm to enhance its applicability in IoT scenarios\cite{gao2013improved,li2013smart,chen2019improved}. Nevertheless, replacing the RETE algorithm is challenging in practice due to the complex implementation of general rule engines, such as Drools\footnote{https://github.com/kiegroup/drools}, which are designed to adapt to various application scenarios.

General rule engines are not the only solution. The open-source IoT platform, Thingsboard\footnote{https://github.com/thingsboard/thingsboard}, deviates from the RETE algorithm in its rule engine implementation and proposes an alternative rule model for streaming data processing. The Thingsboard rule engine offers a more user-friendly interactive model compared to general rule engines and has found wide application IoT scenarios. However, the strong coupling between the components of Thingsboard makes it challenging to apply its rule engine individually in other systems. Moreover, the Thingsboard rule engine does not support the monitoring of data flows from multiple devices within a single rule.

Both the aforementioned primarily applied solutions have potential limitations. General rule engines based on the RETE algorithm are not well-suited for handling continuously changing data sets. The Thingsboard rule engine, while feature-rich, suffers from poor portability.

To address these issues, this paper introduces an adaptable and easily extendable IoT-specific rule engine framework. We defined a basic Domain-Specific Language (DSL) for rule description. This framework can be easily integrated into almost any structured IoT system. It is primarily designed to modify device data flow monitoring rules during the run-time of the IoT system.

We implemented a prototype system of proposed framework. And we validated the theoretical approach through a series of test cases running at the prototype system. After that, we discussed the framework’s effectiveness, scalability, limitations, and application scenarios.

This paper begins by presenting the outcomes and limitations of general rule engines and Thingsboard rule engine, subsequently establishing the research objectives. 

Subsequent chapters propose IoT-specific rule engine framework and the DSL for rule description. The design of rule DSL is presented, followed by an outline of the flow of rule matching function in accordance with its syntax. Then we provided the method for parsing for and scheduling rules.

The latter part of the paper details the implementation of the prototype system, test cases, and results. At last, we made a comprehensive discussion and summary of the proposed framework .

\section{Related Works}
\label{sec:headings}
The RETE algorithm is widely used in general rule engines for pattern matching. Pattern matching means matching data, referred to as fact, with a rule. This algorithm optimizes the computational load by compiling user-defined rules into a network structure know as RETE network. 

The RETE network designed the structure of Alpha Memory and Beta Memory, which stores intermediate results of each pattern matching. However, in scenarios with rapidly changing datasets, such as IoT, the reuse rate of intermediate nodes in RETE networks is low, leading to low cost performance in generating complex RETE networks. To address this limitation, scholars have made improvements to the RETE algorithm in the IoT field\cite{gao2013improved,li2013smart,chen2019improved}. 

Due to the complexity of the RETE algorithm, IoT system developers often choose to use a general rule engine based on RETE, such as Drools, rather than implementing it from scratch. Drools offers a comprehensive implementation of RETE and is widely used in IoT senarios. However, general rule engines like Drools have high structural complexity, making it challenging to replace the RETE algorithm. 

An alternative option is the open-source platform Thingsboard, which provides a dedicated rule engine for IoT device data flow processing. Thingsboard offers a modular rule editing interface with higher usability compared to Drools. Some researchers and developers have chosen to consolidate data visualization and data flow rule management into Thingsboard instead of using a general rule engine\cite{de2018sensor,kadarina2018monitoring}. The rule engine of Thingsboard is implemented as a chain structure, which is known as rule chain.

A rule chain primarily consists of three types of rule nodes:
\begin{itemize}
    \item Filter node for filtering and routing message attributes.
    \item Enrichment node for updating the metadata of incoming messages.
    \item Action node for executing various actions based on the results of incoming message matching.
\end{itemize}
These three categories of nodes corresponding to the Rules, Facts, and Actions of the RETE algorithm. Thingsboard decomposes complete rules into rule nodes and combines these nodes into a rule chain to enable flexible rule configuration.

However, despite offering a comprehensive solution for IoT data analysis and visualization, Thingsboard rule engine has certain limitations. Firstly, it is tightly integrated with other modules of the platform, making it challenging to migrate to alternative systems. Secondly, rules are defined directly on individual devices, preventing users from creating rules that contains multiple devices.

Drools stores data that will be processed in a unified working memory and delegates all rule actions to a single executor. These designs are well-suited for the scenario discussed in this paper. Therefore, the framework presented in this paper intends to adopt these approaches by consolidating all data for rule monitoring into a shared memory and executing all actions uniformly using a separate executor thread.

However, the RETE algorithm, which significantly impacts the performance issues of Drools in IoT data monitoring, will not be utilized in the framework proposed in this paper.

Thingsboard's practice of assigning rules directly to specific devices, which binds a particular data stream from the input platform to a specific process, poses challenges in defining rules for monitoring multiple devices. To address this, the paper suggests importing all monitored data flows into a public cache, allowing unrestricted access to all rules.

Furthermore, all rule nodes in Thingsboard inherit from a common superclass. This inheritance structure leads to a certain level of interdependence between action nodes and condition nodes. To enhance the flexibility of our framework, we proposed a complete separation of the processes of evaluating conditions and executing actions, with only the necessary parameters for action execution being transferred between them.

Thingsboard's rule model implements a corresponding rule node class for each condition and action, which is intuitive and user-friendly. The proposed framework inherits this design pattern. 

\section{IoT Rule Engine Framework}
\label{sec:others}
This paper proposes a framework of IoT rule engine that allows users to customize data flow monitoring and control strategies in run-time of the IoT system. This framework enables users to upload new rules through REST API interfaces and control the activation state of these rules. Each rule specifies the datasource it applies to, the condition it evaluates, and the action to be executed when the rule is matched. The active rule will be converted into a match function that is periodically called.

To establish a common understanding, we introduce two notations: $M$ and $E$. The notation $M$\{K|V\} represents a map with a key K and its corresponding value V. Similarly, $E$\{V1,V2\} defines a structure with data member V1 and V2.

\subsection{DSL for Rule Description}
In order to enable the execution of user-customizable monitoring and control strategies, we defined the syntax of rules in the following manner.
\begin{itemize}
    \item Datasource: \{name\{device\_id,\ device\_type,\ attribute\}\}
    \item Condition: \{expression\ of\ name\}\ or\ \{condition\_type: params\}
    \item Action:\{action\_type:string\ contains\ symbol\ \$name\}
\end{itemize}

The Datasource field specifies the name and attributes of the datasource. The term datasource represents real-time data obtained from a data flow. 'device\_id' serves as a unique identifier for the device within the system. 'device\_type' records the type of the device. 'attribute' refers to a field in the message uploaded by the device, indicating the specific data that needs to be monitored. Multiple datasources can be listed in this field, separated by a semicolon.

The Condition field defines the match condition for the data source. There are two types of conditions: expression condition and functional condition. Expression condition is made by the expression contains datasources. Functional condition specifies condition type with symbols.

The Action field determines the action that will be executed if the rule matches successfully. The \$ symbol can be used to refer to a datasource defined within the same rule.

\begin{figure}[htbp]
    \centering
    \subfigure[Usage of Rule with an Expression Condition]{
    \label{fig:1.sub.1}
    \includegraphics[width=0.74\linewidth]{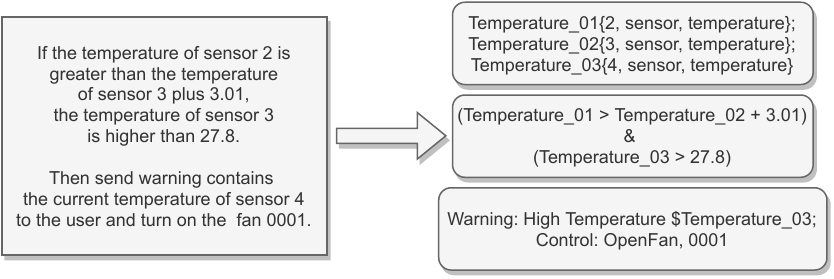}}
    \subfigure[Usage of Functional Condition]{
    \label{fig:1.sub.2}
    \includegraphics[width=0.54\linewidth]{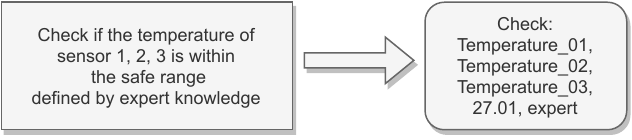}}
    \caption{Example of Rule DSL}
    \label{fig:1}
\end{figure}

Figure \ref{fig:1} provides an example of rule DSL. Figure \ref{fig:1.sub.1} demonstrates a usage of expression condition. Figure \ref{fig:1.sub.2} illustrates a usage of functional condition.

\subsection{Process of Match Function}
First and foremost, it is important to comprehend the functioning of the match function (MF) that is produced by the rule parsing function and embodies our methodology for achieving rule matching during runtime. In corresponding with the three rule statements, the operation of MF can also be conceptualized as consisting of three components: obtaining the most recent data, performing conditional evaluation, and executing the necessary actions.

\subsubsection{Maintaining and Retrieving datasources}

\begin{figure}[htbp]
    \centering
    \includegraphics[width=0.7\linewidth]{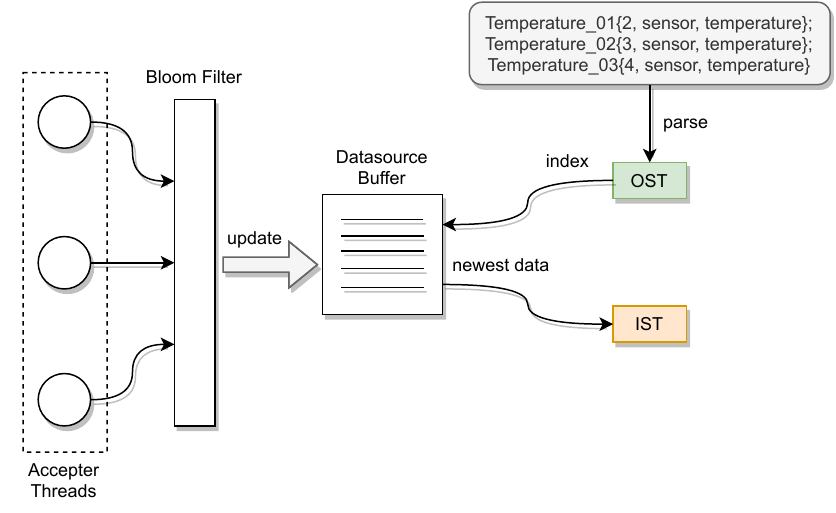}
    \caption{Maintaining and Retrieving Current Data}
    \label{fig:2}
\end{figure}

Figure \ref{fig:2} outlines the logic for maintaining and retrieving up-to-date data.

The rules store the up-to-date values of their data sources in a global map called the data source cache(DSB). When the MF is executed, it retrieves the necessary data for the rule from DSB. The following is the definition of DSB.

\[DSB=M\{Index|E\{Reference,Data\}\}\]

Where Index is defined as follows.

\[Index=E\{device\_id,device\_type,attribute\}\]

'Reference' refers to the quantity of rules that mentions a particular datasource, while 'Data' refers to the up-to-date value of datasource.

MF uses the Datasource field as a reference to retrieve the relevant data from the DSB. In order to facilitate this process, a symbol table is introduces. Since the Datasource field is initially in string format when it is passed into the platform, it cannot be directly used as an index for querying. As a result, during the parsing of MF, the Datasource in string format is converted into another symbol table. This new symbol table uses symbols as indexes and the corresponding Indexes in DSB as values. The symbol table that stores the up-to-date data Index associated with each symbol is referred to as the outer symbol table (OST).

\[OST=M\{Symbol|Index\}\]

The inner symbol table (IST) stores the most up-to-date information about symbols.

\[IST=M\{Symbol|Data\}\]

Therefore, the OST is used to retrieve the up-to-date data from DSB, while the IST is used to receive and store these data. 

It is crucial to emphasize that the DSB should consistently hold the latest data received throughout the system. The datasources within DSB are updated by a set of threads known as the accepter threads (AT), which are responsible for receiving external data. For each attribute in received messages, AT queries DSB to determine if a datasource exists. If exists, it is updated accordingly.

In IoT systems, there exists a notable disparity between the number of message attributes that users are concerned with and the total number of attributes contained within the message. The attributes found in a message typically include upload time, session number, device ID number, device type, communication protocol, environmental data, and other fundamental information. The majority of these attributes pertain to control information, which ensures communication reliability and message legibility, with only a small portion being environmental data of interest to users. As the DSB exists as a global data structure, it necessitates certain synchronization measures to ensure safe progress of concurrent reads, resulting in concurrency overhead during the query process. To address these considerations, a Bloom Filter\cite{bloom1970space} is used to reduce the frequency of AT accessing DSB.

\subsubsection{Calculation of Specified Condition with datasources}
\begin{figure}[ht]
    \centering
    \includegraphics[width=0.42\linewidth]{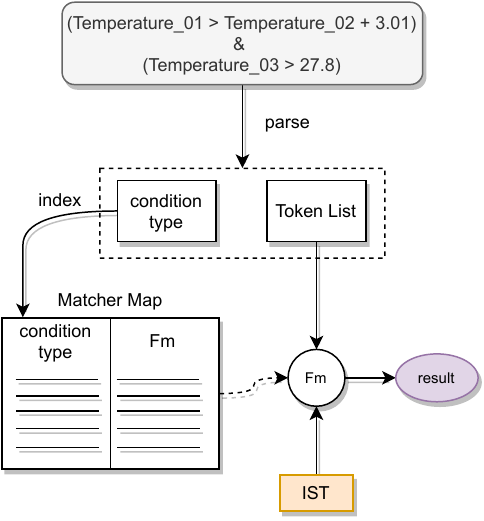}
    \caption{Calculating the Rule Matching Results}
    \label{fig:3}
\end{figure}

Figure \ref{fig:3} depicts the logic for computing the results of rule matching.

After retrieving the latest data and creating the IST, we use data contains in it to perform the computation as specified in the rule. Subsequently, we analyze the Condition field to determine the type identifier of the condition and the list of Token. The term Token is defined as follows.

\[Token=E\{type,value,real\_num\}\]

The term 'type' means the classification of the Token, which can encompass variable types, bracket types, operator types, numeric types, or string types. The term 'value' denotes the string representation of Token. The term 'real\_num' signifies the present value of the string. It is noteworthy that solely variable type and numeric type Token will possess a populated 'real\_num' field.

Subsequently, we use the condition type identifier to index Matcher Map to obtain the pattern matching function ($F_m$) that corresponds to this condition type.

\[Matcher Map=M\{condition\_type|F_m\}\]

We proceed by inputting the list of ISTs and Token into the function $F_m$ to obtain the execution result.

For functional conditions, the input parameter list of $F_m$ contains only two types of Tokens. One of them is variable Token, which can be used to query for a value of the IST. Another type is string Token, which will be parsed by $F_m$. In order to unify the handling of the two conditions, a condition type symbol will be auto assigned as index during parsing expression condition. The computation process for $F_m$ of expression condition involves replacing the variable Token in the list with the latest data, and then using a stack-based expression evaluation algorithm to calculate its value. This entails sequentially sending the Token list of the succeeding expression order to a stack, where numerical Tokens are directly placed into it. Operator Tokens will extract the top two elements from the stack, perform the corresponding operation, and push the result back into the stack until the final result is achieved.

\subsubsection{Execution of Specified Actions}

\begin{figure}[htbp]
    \centering
    \includegraphics[width=0.86\linewidth]{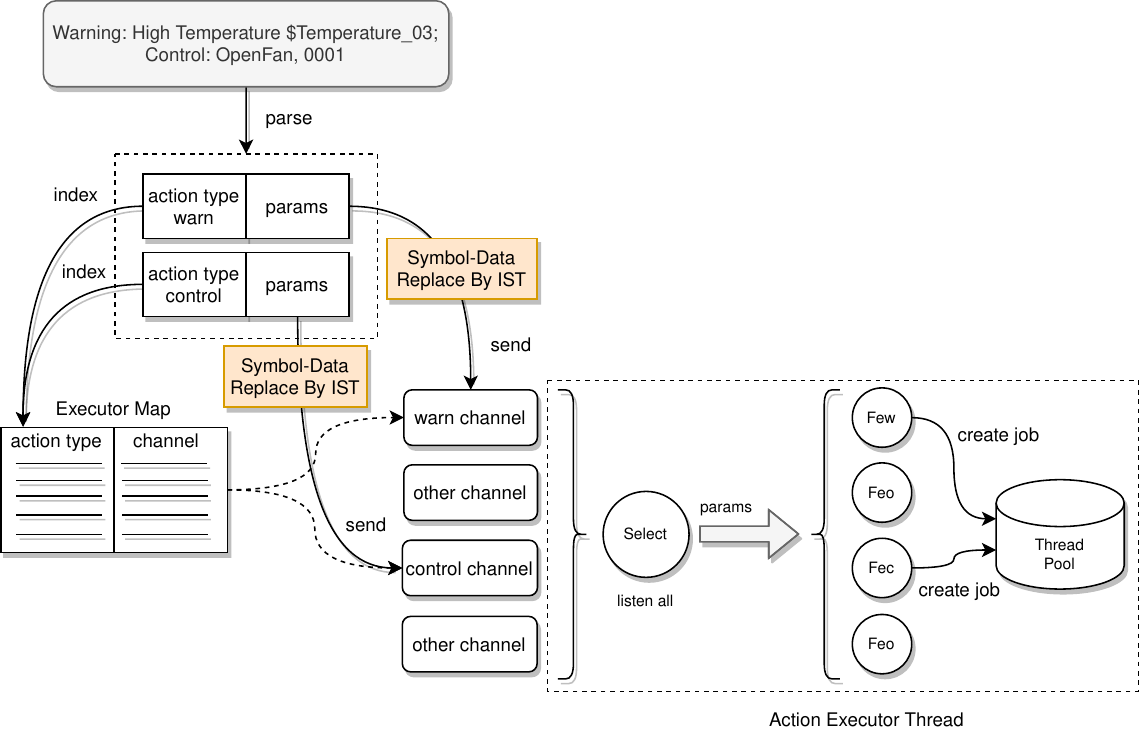}
    \caption{Executing Actions}
    \label{fig:4}
\end{figure}

Figure \ref{fig:4} illustrates the logic for executing actions within the rule.

After applying the conditional type to index $F_m$, the list of tokens is processed with the IST. This process means matching the most recent data with the rules. The actions specified in the Action field are executed when a match is found.

First, it is assumed that each action is executed by a corresponding execution function ($F_e$). This assumption is valid because even complex action logic can be encapsulated into a callable function. The function takes all the necessary parameters for action execution, which enhances code organization and maintainability, facilitating future development.

Under this assumption, the calling mode of $F_e$ is further analyzed. Directly calling $F_e$ by MF is not advisable as it binds the execution results of multiple $F_e$s. If an error occurs during execution, all subsequent $F_e$s will be affected. Additionally, the execution order of multiple $F_e$s at the same level may vary depending on the order in which they are called by the main function.

One possible solution is to create a thread for each $F_e$ to handle the execution. However, this approach has two drawbacks. Firstly, creating numerous threads can disrupt normal system operation, especially when there are a large number of rules. Secondly, there is no clear separation between condition matching and action execution, which hinders system modularization and distribution.

To address these issues, each $F_e$ is assigned a channel for transferring the parameter list. A channel can be seen as a shared memory area implemented using the consumer-producer model. In the Executor Map, channels are organized and categorized based on their respective action types.

\[Executor Map=M\{action\_type|channel\}\]

In the proposed approach, the parameter list is sent to the channel in string format by the MF after a successful match. It is important to note that any datasources indicated by the \$ symbols in the string are replaced with numerical values before being sent. The channels are all listened to by a single thread, referred to as the action executor (AE). AE then forwards the received parameter lists to their corresponding $F_e$s. These $F_e$s are executed by a public threadpool. This design aims at limit the system resources that could be occupied by action execution, reduce the overhead associated with creating and destroying multiple threads, and effectively separate the processes of condition matching and action execution into two independently running modules.

\subsection{Translation Process from Rule Text to Runtime Structure}

Once the process of executing MF is comprehended, the subsequent inquiry concerns the production of MF from the rules. Specifically, it entails understanding how the rule parse function modifies the current system while interpreting the rule text and how it constructs the corresponding MF.

\subsubsection{Parsing Datasource and Creating OST}
The first step of creating a MF involves parsing the Datasource field and constructing the OST.

\subsubsection{Condition Parsing, Categorization of Matching Conditions, and Token List Generation}
After the initial step, the Condition field is evaluated based on its category. Depending on the category, different parse functions are called to generate the Token list. For conditions of the expression type, the string expression will be initially parsed into a mid-order Token list. This mid-order Token list is then converted into a post-order Token list, which allows the expression evaluation algorithm, $F_m$, based on the Token Stack, to perform calculations. For conditions of the functional type, they will be directly parsed into a Token list.

\subsubsection{Action Parsing and Action List Generation}
Parsing the Action field is a straightforward process. Each action type identifier and its corresponding parameters will be parsed as a structure named Action.

\[Action=E\{action\_type,params\} \]

These structures collectively constitute the Action List (AL).

\subsubsection{Creation of Match Function}
Matcher generator function returns the MF associated with the rule by using the OST, condition type, Token list, and AL as input arguments. The flow of the matcher generator function is to create and return a function that does not require any inputs or outputs. The generated function runs as follows.

\begin{enumerate}
    \item The OST is used to query DSB to obtain the latest data of the symbols and construct the IST from it.
    \item Subsequently, condition type is used to index and call the $F_m$, with IST and Token list as input parameters. The return value of this call serves as a condition for determining whether the pattern matching is successful or not.
    \item If the match is found, the AL is iterated, an IST-based symbolic-value substitution of the datasource is performed on the parameter list, and it is forwarded to the corresponding channel after substitution.
\end{enumerate}

\subsection{Rule Scheduling}

The corresponding MF is specifically designed for the execution of rules. However, the framework also provides the capability to include rules that do not need to be executed immediately, allowing users to register them in the system. In order to facilitate scheduling, the framework defines four states for these rules: undefined, inactive, scheduled, and active. Additionally, the framework outlines the transition relationship between these states.

Scheduling logic of rules will be implemented with the assistance of an active rule map, a scheduled rule map, and a Rule Database (RDB).

\begin{figure}[htbp]
    \centering
    \includegraphics[width=0.58\linewidth]{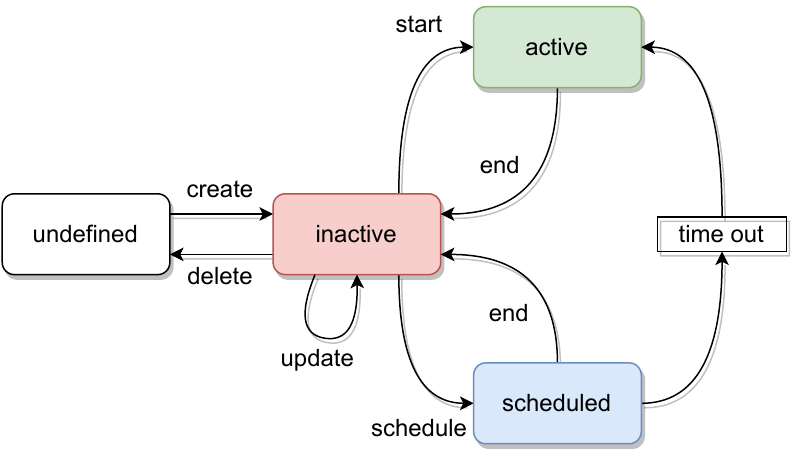}
    \caption{Scheduling Logic of Rules}
    \label{fig:5}
\end{figure}

Figure \ref{fig:5} elucidates the connection between rule state transitions and scheduling interfaces.

\subsubsection{Create Rules Interface}
The rule is created and be saved to the RDB with the rule state initialize as 'inactive'.
\subsubsection{Start Rules Interface}

In order to match up-to-date data with rules in active state, a process known as the matcher thread (MT) runs continuously to periodically invoke MF. We found that the Cron\footnote{https://pkg.go.dev/github.com/robfig/cron} timed task framework is suitable for the requirements of our application. Therefore, we create a global Cron object and use it to execute all MF. The implementation is straightforward. We simply provide the execution period and the generated MF to the Cron object.

To ensure the latest data will be accessed and updated by AT, rules stored in the RDB are extracted and the Datasource is preparsed. Corresponding entries are then added to DSB, which increases the reference count of the target datasource or creates the datasource if it does not already exist. After this step, the rule text is parsed into MF. MF is then added to the Cron. Corresponding Cron entry ID is saved in the active rule map, with the Rule Table ID (RID) serving as the index. Finally, the rule status is updated to 'active'.

\begin{figure}[htbp]
    \centering
    \includegraphics[width=0.7\linewidth]{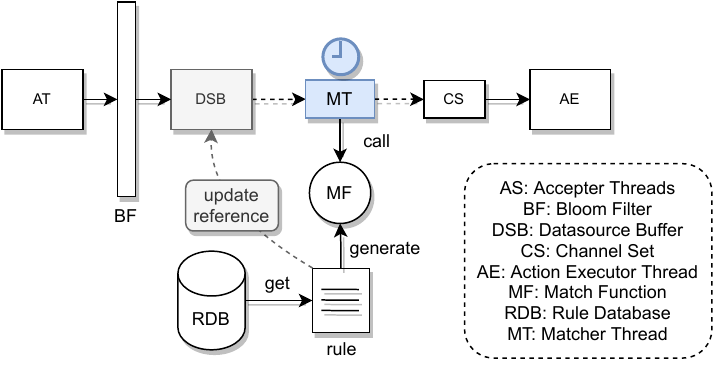}
    \caption{The Process of Activating a Rule}
    \label{fig:6}
\end{figure}

Figure \ref{fig:6} illustrates the process of activating a rule.

\subsubsection{Schedule Rules Interface}

A new timer, denoted as 'Timer,' is added into the scheduled rule map, with indexing based on the RID. Upon reaching its designated time, 'Timer' will call the 'Start Rules' interface and subsequently eliminate itself from the scheduled rule map. In the mean time, rule status will change to 'scheduled'. After a while, the rule status will undergo a transition from 'scheduled' to 'active' when timeout.

\subsubsection{Update Rules Interface}
Rules in RDB with a status of 'inactive' are updated. The rule status does not need to be updated.

\subsubsection{End Rules Interface}

The process of ending rules involves removing the corresponding entry in DSB, which decreases the reference count of the target datasource. If this reduction results in a count of zero, the entire datasource is deleted.

If the target rule has a status of 'scheduled' in the RDB, its RID is used to retrieve and delete entries from the scheduled rule map. Additionally, the corresponding Timer is halted.

On the other hand, if the target rule has a status of 'active' in the RDB, its RID is utilized as an index to fetch and delete entries from the active rule map. The fetched object entry is then used to remove the corresponding task from the Cron.

Ultimately, the status of the rule is modified to 'inactive'.

\subsubsection{Delete Rules Interface}
Rule with a status of 'inactive' in the RDB will be deleted.

\section{Evaluation}
A prototype system has been developed to evaluate the efficacy of the framework proposed in this paper.

\subsection{Prototype System with Proposed Rule Engine Framework}
We commence our analysis by examining the theoretical application scenarios of IoT rule engine framework. Subsequently, we proceed to develop a prototype system for the purpose of testing framework.

\subsubsection{Theoretical Application Model}

The data generated by devices is received and processed through AT. The data is then stored in MongoDB for long-term storage and historical data analysis. Additionally, the DSB is continuously updated with real-time data. To send control commands to the devices, the MQTT Broker is utilized, where the devices only need to subscribe to a specific MQTT topic for this purpose. Furthermore, our primary service database, MySQL, serves as RDB in this scenario.

\begin{figure}[htbp]
    \centering
    \includegraphics[width=0.72\linewidth]{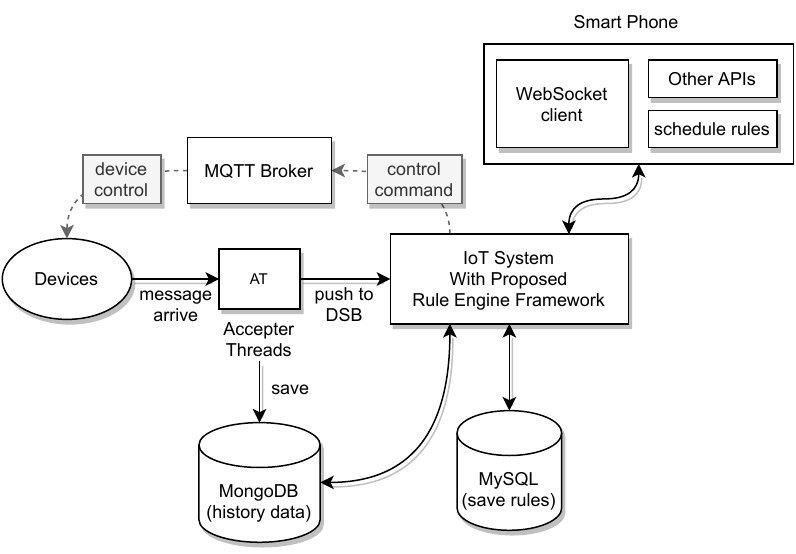}
    \caption{Theoretical Application Model}
    \label{fig:7}
\end{figure}

Figure \ref{fig:7} presents the general application model.

\subsubsection{Prototype System}
The prototype system was constructed with the intention of fulfilling the following objectives.
\begin{itemize}
    \item To simulate data flow send by IoT devices, we use a Python script contains a TCP client, which transmits data to the server. Additionally, we implemented an AT , which contains TCP server, to receive data.
    \item To simulate the process of device command reception, we use a MQTT client to subscribe to a specific topic.
    \item To simulate the process of a user-side smartphone receiving information pushed from the server, we use a WebSocket client to establish a connection with the prototype system.
    \item For testing the functionality of our framework and simulating user actions, we utilized the Swagger UI web page to call the REST API interface provided by the system.
\end{itemize}

In our \href{https://github.com/ChenKen9869/DCA-IOT-system}{prototype system\footnote{https://github.com/ChenKen9869/DCA-IOT-system}}, we have implemented $F_e$ for pushing MQTT messages and $F_e$ for delivering messages to WebSocket clients. Furthermore, we have developed a collection of other interfaces, which are also essential for the system's functionality. These interfaces include WebSocket connection services.

\begin{figure}[htbp]
    \centering
    \includegraphics[width=0.76\linewidth]{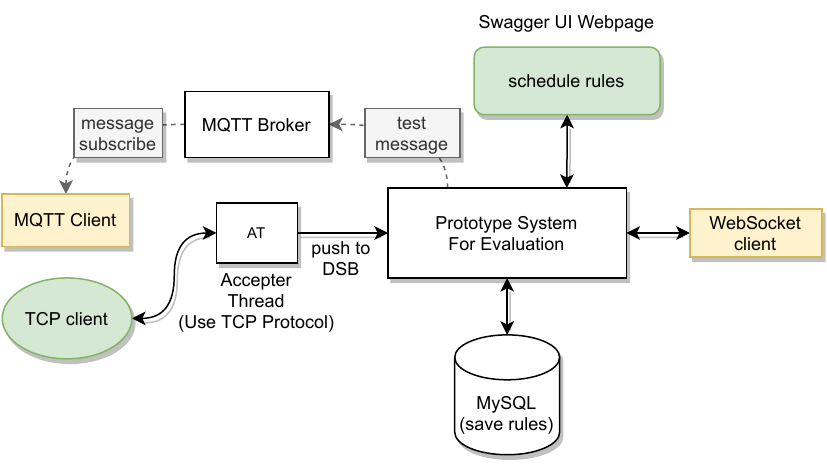}
    \caption{Prototype System for Testing}
    \label{fig:8}
\end{figure}

Figure \ref{fig:8} illustrates the prototype system used for testing purposes.

\subsection{Test Cases and Results}
The following section presents the test cases used in the prototype system testing. We deployed the prototype system on the Ubuntu 22.04 operating system. Each of these test cases yielded the anticipated outcomes, thereby affirmed the effective validation of our proposed framework.

\textbf{Test-1: Validation of Single-Device Rule with Expression Condition and Multiple Actions}

\textbf{Rule 1}:

\begin{itemize}
    \item Datasource: tem\{1, Portable, temperature\} 
    \item Condition: tem > 22.1
    \item Action: WebSocket: 1,rule Matched, temperature is \$tem!;Mqtt: localhost, 1883, admin, emqx@123456, test, control temperature
\end{itemize}

In order to evaluate the effectiveness of the rule, we performed an experiment using a Python script to transmit a range of messages to the server. It was observed that only a portion of these messages which satisfied the specified conditions were processed. These results obtained from this experiment aligned with our initial expectations.

\begin{enumerate}
    \item MQTT client received expected message only if temperature attribute of message is greater than 22.1.
    \item WebSocket client received expected message only if temperature attribute of message is greater than 22.1.
\end{enumerate}

The testing methodologies for the following rules are similar. Both of them involve modifying the message body in response to specific conditions and then verifying if the MQTT client and WebSocket client receive the expected messages.

\textbf{Test-2: Validation of Multi-Device Linkage Rules with Expression Condition and Multiple Actions}

\textbf{Rule 2}:

\begin{itemize}
    \item Datasource: tem\_1\{1, Portable, temperature\}; tem\_2\{1, Fixed, temperature\}
    \item Condition:  (tem\_2 > 25.3) \& (tenm\_1 > tem\_2 +3 )
    \item Action: WebSocket: 1,rule Matched, temperature is \$tem\_2 and \$tem\_1!;Mqtt: localhost, 1883, admin, emqx@123456, command, open fan
\end{itemize}

\textbf{Test-3: Validation of Functional Condition through Multiple Actions}

\textbf{Rule 3}:

\begin{itemize}
    \item Datasource: longitude\{3, Portable, longitude\}; latitude\{3, Portable, latitude\}
    \item Condition: PointSurface: longitude, latitude, 1xx.40xxx2, 3x.92xx55, xx6.4xx70x, xx.89xx55, 1xx.40xx92, x9.8xx353, xxx.38xx46, xx.89x365
    \item Action: WebSocket: 1,rule Matched, position is \$longitude \$longitude!;Mqtt: localhost, 1883, admin, emqx@123456, command, find device
\end{itemize}

The PointSurface functional condition of Rule 3 is used to verify the inclusion of a point within a designated surface. This rule is utilized to establish whether the target device is situated within a predetermined range. From the third parameter onwards, the longitude and latitude coordinates of the points that define the polygonal boundary of the specified range are populated in a sequential manner. In order to safeguard privacy, certain numerical values in the test case have been obscured, with the use of 'x' as a replacement.

\textbf{Test-4: Simultaneous Execution of Multiple Verification Rules}

To ensure the successful attainment of desired outcomes when executing multiple rules concurrently, we commenced testing by starting Rule 1, Rule 2, and Rule 3 concurrently. And also using REST API interfaces to schedule them. During this phase of concurrent testing, the predetermined relationships and conditions for state transitions operated as expected. Subsequently, we extended our testing to encompass additional rules, and despite the higher complexity, we consistently obtained the expected outcomes.

\textbf{Test-5: Testing of the Average Memory consumption of a Single Rule}

\begin{figure}[htbp]
    \centering
    \includegraphics[width=1\linewidth]{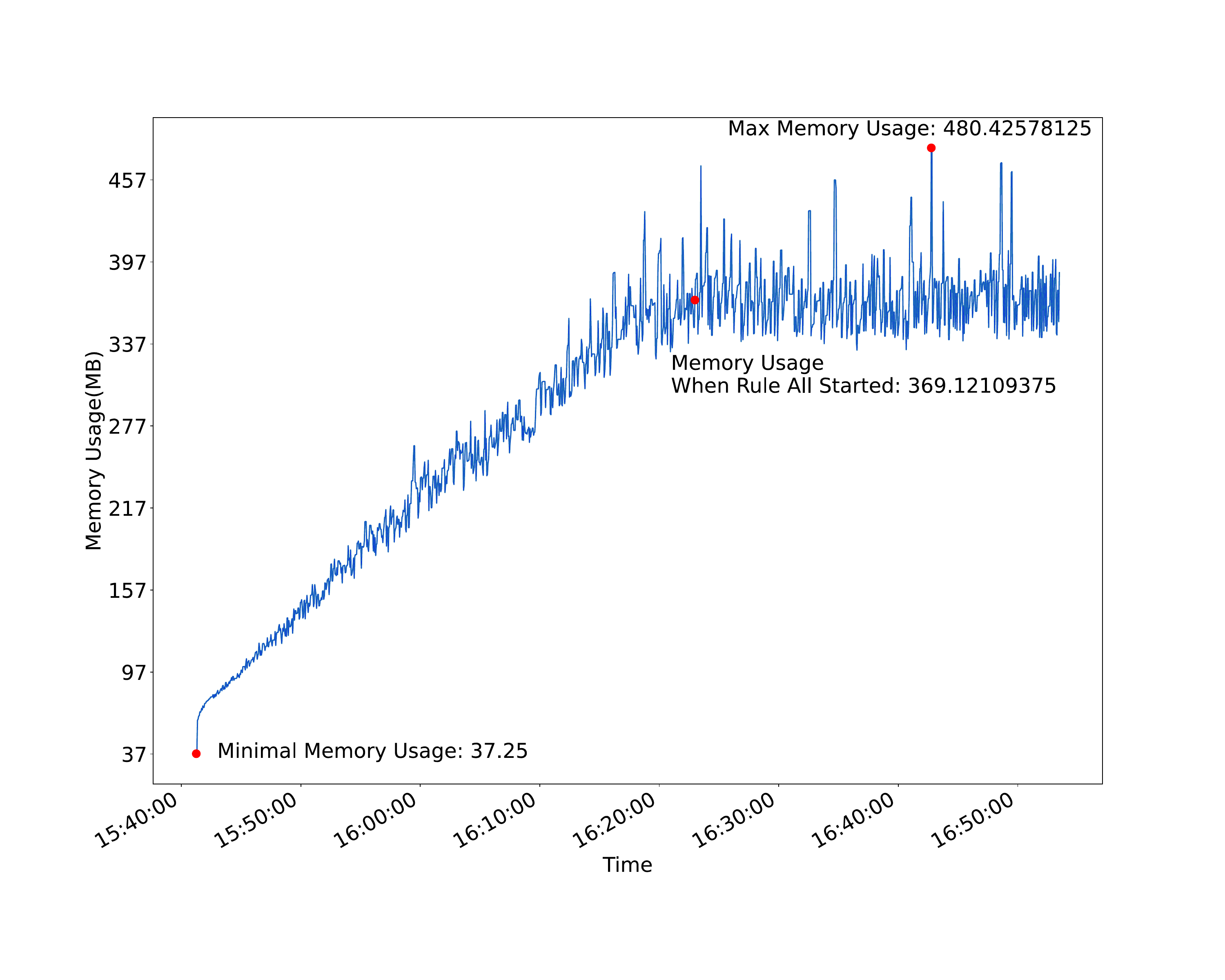}
    \caption{Memory Usage of Prototype System}
    \label{fig:9}
\end{figure}

We tested the average memory usage of a single rule by running 100,000 rules concurrently in the system. All testing rules have the same definition with Rule3. All rules are executed at a 5-second interval.
Figure \ref{fig:9} illustrates the relationship between system uptime and memory usage.

Before we started the concurrent testing, the prototype system had a memory usage of 37.25MB. Then we started to create and run the rules with 20 threads, each responsible for 5000 rules. When the time reached 16:22:59 and all the rules were created, the system memory usage was about 369.12MB. We didn't create any new rules for nearly 30 minutes, and the system's memory footprint fluctuated around 369.12MB, peaking at around 480.43MB. 

We used the peak memory usage of 480.43MB when all rules had started, minus the memory usage of 37.25MB when no rules were started, and then divided by the total number of rules 100,000 to get an average memory usage of about 4.54KB per rule. This result is well lower than the minimum average memory usage of about 300KB for the RETE algorithm and the RETE improved algorithm family\cite{chen2019improved}.

\section{Discussion}

After verified the theoretical correctness of our framework, we will now engage in a comprehensive examination of its validity, challenges, potential methods for addressing limitations, and the usage scenarios.

\subsection{Effectiveness}
In light of the research background and research objectives previously discussed, we assess the effectiveness of the proposed framework by considering both theoretical perspectives and empirical findings. Through our experimental verification, we have confirmed the validity of the suggested rule engine. Our subsequent discussion primarily focus on the applicability, ease of use, and potential for expansion of our framework.

\subsubsection{Easily Integrated with almost any IoT system}
The proposed framework can be easily integrated into the existing mainstream structures of IoT system.
\begin{itemize}
    \item \textbf{Resource Shortage System}: The suggested framework does not impose substantial memory overhead, in contrast to the RETE algorithm, rendering it appropriate for systems with limited resources. Furthermore, the comprehensive implementation of the framework can be accomplished as long as the deployed system has multithreading capabilities. In situations where the underlying hardware operates exclusively in a single-threaded mode, it is feasible to integrate the framework within the higher-level software of the system.
    \item \textbf{Distributed System}: Maintaining a low level of coupling between individual components is of utmost importance in order to ensure the smooth functioning of a distributed system. The necessary modifications to accommodate a distributed system are not significant. For example, in the case of the DSB, it is adequate to encapsulate it within a separate microservice and establish external interfaces for updating, deleting, and querying data sources. Likewise, for components such as the AE that require substantial resources, we can manage them within separate microservices and provide dedicated interfaces for receiving lists of action parameters.
    \item \textbf{System dependent on third-party services}: In an IoT system that relies on a third-party service, certain devices within the system do not directly send messages to the main server. Instead, they send messages to a server managed by the manufacturer of the device. This server then responds to HTTP data requests from users of the device. In this particular situation, our main objective is to create an AT that regularly queries the third-party service to obtain the data and subsequently transfers it to DSB.

\end{itemize}

\subsubsection{User-friendly}
The rule syntax of Drools is based on the Java language, which may pose a degree of complexity for users. In contrast, Thingsboard rule engine utilizes a modular user interface and JavaScript language for rule creation, rendering it relatively more accessible to users compared to Drools. The framework proposed in this paper adopts a specifically designed DSL for rule deployment. This DSL consists of only three statements with a simple syntax that can be readily comprehended and mastered by users without extensive computer expertise. Moreover, scheduling rules through REST APIs within the framework is also user-friendly.

\subsubsection{Effortlessly Scalability}

The framework discussed in this paper have the room for improvement of providing users with the ability to define their own data processing procedures using a general-purpose programming language. Thingsboard incorporates a JavaScript interpreter service to enable customizing rule nodes. Our framework can also support customizing Condition and Action by wrapping interpreters for script languages, such as JavaScript or Python, as $F_m$ or $F_e$.

Primary components, MF, AE, and AT, demonstrate low coupling. This is evident as there are no direct calls between these components. Instead, their dependencies are established through indexing or intermediate caching mechanisms. As a result, extending $F_m$, $F_e$, and AT is a straightforward process.

The extensibility of the rule DSL in the framework is enhanced by incorporating design principles from general-purpose programming languages. For example, the rule DSL is able to expand to allow for the definition of constants. These constants could be parsed and stored in a constant table along with their symbols and data values. The constant table can then be queried for future use. Additionally, rule chaining can be implemented in the framework as well. Rule chaining means automatically triggering the next rule upon the successful matching of the current rule. If the preceding rule fails, the subsequent rule will not be triggered. This is achieved by introducing a new $F_e$, which modifies a specific cell of DSB to a predefined value. Then the preceding rule uses such $F_e$ to update DSB, while the subsequent rule take corresponding DSB cell as its datasource.

\subsection{Challenges and Feasible Countermeasures}
In the current framework, pattern matching is triggered periodically by the Cron. The trigger condition is based on the timing of Cron's execution interval. When triggered, data corresponding to the rule is retrieved from DSB. However, the framework lacks the ability to check if the retrieved data has been previously validated or to detect message omissions. This could lead to the processing of duplicate messages if the datasource in the DSB has not been updated when pattern matching occurs. Additionally, if multiple datasources are updated quickly within the pattern matching execution interval, only the latest one will be used for pattern matching.

One possible solution is to keep a record of message upload frequencies for each device type and align the pattern matching execution period with this frequency. However, network congestion or downtime of IoT devices can still cause delays in updating DSB, potentially resulting in message duplicates. Network congestion can also lead to delayed message arrivals, causing consecutive messages arrive and message omissions within a single execution cycle. While setting an appropriate pattern matching execution cycle can mitigate message omissions and duplicates, alternative trigger conditions may be necessary in certain special scenarios.

To prevent duplicate data processing, a session field can be introduced in both DSB and OST to track duplicate data. Before each matching execution, the session of the data source in the OST is compared to the session in the data source cache. If the OST session is smaller, the session in the OST is updated, and matching proceeds. This approach helps avoid message duplication and detects message omissions, but it does not guarantee against message omissions.

For scenarios where both preventing duplicate data processing and absolute avoidance of missing data are crucial, a push trigger condition can be used. In this approach, DSB proactively pushes new data to the MT by a channel. Instead of storing the actual data values in DSB, a map is maintained in DSB with the RID as the key and the channel as the value. When updating a specific datasource, DSB sends the same message to all channels in the map. Additionally, a continuous loop is used to invoke MF indefinitely, rather than depending on the Cron. MF listens to each datasource channel using the rule ID and the OST. When data arrives in a channel, it populates IST and initiates pattern matching as soon as IST is filled. In this setup, reference counting in the DSB is no longer necessary as the map capacity represents the number of rules requesting data from the source.

When implementing push-style trigger conditions as described above, it is advisable to avoid creating rules for multi-device processing, especially when the message arrival frequencies vary significantly between devices. This is because building the IST can become slower, especially if data from one symbol arrives significantly earlier than data from another symbol. If several batches of new messages arrive during this time, the threads responsible for sending data to the channel in DSB may become blocked for an extended period, impacting the update process for other data sources. In systems using such push-type trigger conditions, controlling the execution period may not be suitable as it could block the DSB update thread if data is not fetched within the execution interval.

\subsection{Usage Scenarios}
Based on the aforementioned analysis of effectiveness and limitations, it is apparent that the framework possesses a broad range of potential applications. However, it is necessary to mention that the framework may not be suitable for situations that necessitate strict real-time requirements. This is due to the fact that device control relies on periodic pattern matching executions rather than continuous data flow monitoring, which could potentially result in delays in issuing control instructions.

The rule engine framework proposed in this paper is well-suited for IoT scenarios such as home automation, agricultural applications, and personal wearable devices. However, it may not exhibit optimal performance in scenarios that require precise real-time control.

\section{Conclusion}
This paper presents a IoT-specific rule engine framework. The framework contains a user-friendly DSL and enables the seamless modification of data flow monitoring and control strategies without causing disruptions to IoT system operations. 

The framework is characterized by its high scalability and versatility, making it suitable for a wide range of IoT scenarios, particularly those that do not require strict real-time device control. Proposed framework stands out in addressing challenges related to monitoring IoT data flows by minimizing additional memory consumption compared to rule engines based on the RETE algorithm. Additionally, it offers improved flexibility and user-friendliness compared to Thingsboard rule engine.

In future researches, our study will focus on enhancing the robustness of proposed framework. This may involve allowing users to assign priorities to pattern matching or action execution, or detecting semantic discrepancies that may occur when multiple rules are concurrently utilized.

\bibliographystyle{unsrt}
\bibliography{references.bib}

\end{document}